%% file: main.tex
\documentclass[sigconf]{acmart}
\usepackage{textcomp}
\usepackage{graphicx}
\usepackage{booktabs}
\usepackage{multirow}
\usepackage{makecell}
\usepackage{subcaption}
\AtBeginDocument{%
  }

\acmISBN{978-1-4503-XXXX-X/18/06}

\begin{document}

\title{\underline{L}ocation \underline{i}s \underline{K}ey: Leveraging Large Language Model for Functional Bug Localization in Verilog}\thanks{Corresponding authors: Zhe Jiang (101013615@seu.edu.cn).}


\author{Bingkun Yao}
\affiliation{%
  \institution{City University of Hong Kong}
  \city{Hong Kong SAR}
  \country{China}}
\email{bingkun.yao@cityu.edu.hk}

\author{Ning Wang}
\affiliation{%
  \institution{City University of Hong Kong}
  \city{Hong Kong SAR}
  \country{China}
}
\email{nwang227-c@my.cityu.edu.hk}

\author{Jie Zhou}
\affiliation{%
	\institution{Southeast University}
	\city{Nanjing}
	\country{China}
}
\email{zhoujie\_0428@163.com}

\author{Xi Wang}
\affiliation{%
	\institution{Southeast University}
	\city{Nanjing}
	\country{China}}
\email{xi.wang@seu.edu.cn}

\author{Hong Gao}
\affiliation{%
	\institution{Zhejiang Normal University}
	\city{Jinhua}
	\country{China}}
\email{honggao@zjnu.edu.cn}

\author{Zhe Jiang*}
\affiliation{%
	\institution{Southeast University}
	\city{Nanjing}
	\country{China}}
\email{101013615@seu.edu.cn}

\author{Nan Guan}
\affiliation{%
	\institution{City University of Hong Kong}
	\city{Hong Kong SAR}
	\country{China}}
\email{nanguan@cityu.edu.hk}

\renewcommand{\shortauthors}{Yao et al.}
\def\cu#1{{\bfseries #1}}
\def\xi#1{\textit{#1}}
\def\jue#1{$\left|#1\right|$}
\def\juen#1{\left|#1\right|}

\newcommand{\chanmc}[1]{\textbf{{\color{blue}#1}}\xspace}
\newcommand{\bk}[1]{{\color{black}#1}\xspace}
\newcommand{\mody}[1]{{\color{red}#1}\xspace}
\newcommand{\bkmodify}[1]{\textbf{{\color{purple}#1}}\xspace}
\newcommand{\bkreply}[1]{\textbf{{\color{brown}#1}}\xspace}
\newtheorem{Def}{Definition}
\newtheorem{The}{Theorem}
\newtheorem{Cor}{Corollary}
\newtheorem{Lem}{Lemma}
\begin{abstract}
  Bug localization in Verilog code is a crucial and time-consuming task during the verification of hardware design. Since introduction, Large Language Models (LLMs) have showed their strong programming capabilities. However, no work has yet considered using LLMs for bug localization in Verilog code. This paper presents \underline{L}ocation-\underline{i}s-\underline{K}ey, an opensource LLM solution to locate functional errors in Verilog snippets. LiK achieves high localization accuracy, with a \xi{pass@1} localization accuracy of 93.3\% on our test dataset based on RTLLM, surpassing GPT-4’s 77.9\% and comparable to Claude-3.5's 90.8\%. Additionally, the bug location obtained by LiK significantly improves GPT-3.5's bug repair efficiency (Functional \xi{pass@1} increased from 40.39\% to 58.92\%), highlighting the importance of bug localization in LLM-based Verilog debugging. Compared to existing methods, LiK only requires the design specification and the erroneous code snippet, without the need for testbenches, assertions, or any other EDA tools. This research demonstrates the feasibility of using LLMs for Verilog error localization, thus providing a new direction for automatic Verilog code debugging.
\end{abstract}




\received{20 February 2007}
\received[revised]{12 March 2009}
\received[accepted]{5 June 2009}

\maketitle
\section{Introduction}
\input{intro}

\section{Method}
\input{method}

\section{Evaluation} \label{sec:exp}
\input{evaluation}
\section{Discussion} \label{sec:discuss}
\input{discussion}
\section{Conclusion \& Future Work}
\input{conclusion}


\bibliographystyle{ACM-Reference-Format}
\bibliography{sample-base}


\end{document}

%% file: intro.tex
Verilog is an important hardware description language. In the verification of Verilog code design, locating bugs is an crucial and time-consuming task. Thus, automatic bug localization in Verilog code is of great significance and has attracted research attention in recent years \cite{ICCD22-1,FCS24,VLSID16,TCAD22,ICCD22-2,Cirfix,TACO1,DATE24,Strider}. However, existing methods generally require pre-defined components (e.g., test cases, assertions and templates), which need to be carefully written by experts with extensive professional knowledge. This prolongs the debugging cycle of hardware design.

In recent years, Large Lauguage Models (LLMs) have demonstrated their powerful coding capabilities and many LLMs tailored for programming tasks have been proposed \cite{TSE-24,SP-23}. Some works have already considered using LLMs to locate bugs in software languages \cite{FSE-24-2,ICSE-24-3,Arxiv-24,LLM4code-24}. Particularly, by leveraging the code understanding capabilities of LLMs, their method could locate bugs with high precision using only the design description and the target code snippet, without relying on any test cases, expert knowledge, or any code analysis tools. For example, on a C-language benchmark \xi{Devign} \cite{Devign}, \xi{LLMAO} \cite{ICSE-24-3} pretrained from open-source coding LLM \xi{CodeGen-16B} achieved a top-5 localization accuracy of 60.3\%, where top-5 refers to the probability that the buggy line is among the five most likely erroneous lines identified. This inspires us to use LLM for the bug localization problem in Verilog code, which has not been addressed in the literature. Existing works related to Verilog bug localization using LLMs could be divided into the following two categories:

(1) The first category \cite{RTLCoder,Phi-1.5B,VeriGen,DATE-23-2,Veriseek} focused on open-source Verilog-specific LLMs. However, their major concern is Verilog code generation, and their training data is not for locating bugs. Thus, these LLMs are not suitable for the problem of bug localization in Verilog codes. 
\begin{figure}[t]
	\centering
	\includegraphics[width=8cm]{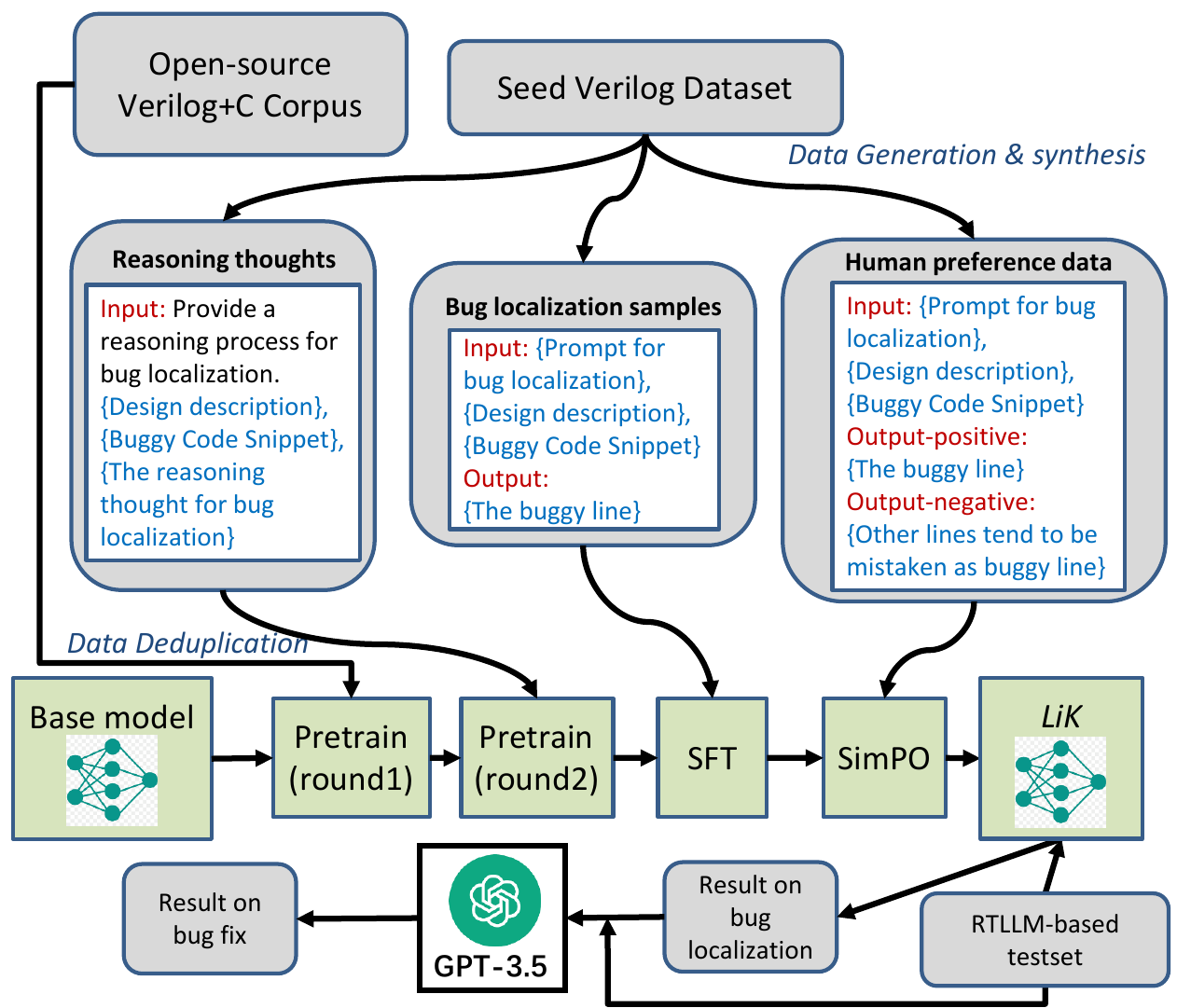}
	\caption{The whole workflow of our research.}
	\label{fig:workflow}
\end{figure}
Generally speaking, to enable a LLM to perform bug localization, it is necessary to use high-quality data related to Verilog bug localization/debugging for model training. Unfortunately, such data is very limited in the open-source community.

(2) The second category \cite{TIFS-24,arxiv-24-2,hdldebugger,MEIC,RTLFixer} investigated the framework for Verilog bug repair, which includes several LLMs as agents and each agent completes one intermediate step. They do not consider model training and typically apply closed-source commercial LLMs (e.g., GPT for \cite{MEIC}, closed-source models\footnote{In the rest of the paper, the terms "LLM" and "model" are used interchangeably.} for \cite{hdldebugger}) as agents. Nevertheless, using commercial LLMs may lead to security and privacy risks brought by code leakage. Besides, given the insular nature of hardware industry, solutions based on open-source models will likely to be more beneficial to the research community.

To address the above issues, this paper proposes an open-source LLM for Verilog bug localization. Particularly, we focus on \xi{functional} bugs, which may lead to incorrect output waveforms without triggering synthesis errors, making them more difficult to be located. (Another type of bugs is syntax error, for which the synthesizer could often directly provide the buggy line.) Specifically, our contributions are as follows:

(1) We propose a LLM named \underline{L}ocation \underline{i}s \underline{K}ey (LiK) to locate functional bugs in Verilog code snippets. LiK is trained through two rounds of continual pre-training (PT), supervised fine-tuning (SFT), and reinforcement learning with human feedback (RLHF). Particularly, in RTHF stage, we leverage human preference data to further reduce the errors made by the model through SimPO algorithm \cite{SimPO}, which has not been considered in existing works on Verilog coding using LLMs.

(2) Experimental results verify the efficiency of LiK, achieving a \xi{pass@1} localization accuracy of 93.3\% on the RTLLM-based \cite{RTLLM} test set, outperforming all current open-source models specially designed for Verilog coding, surpassing GPT-4-Turbo's 77.9\% and comparable to Claude-3.5-Sonnet's 90.8\%. Additionally, we observe that by using the localization results of LiK, the error correction efficiency of GPT-3.5 is significantly improved (Functional \xi{pass@1} increases from 40.39\% to 58.92\%), demonstrating the importance of bug localization in LLM-based Verilog bug repair.

(3) To the best of our knowledge, LiK is the first attempt to use LLM for functional bug localization in Verilog. Our method only requires design descriptions and target code snippets, without the need for other components written by experts and interaction with any EDA tools. This reveals that using LLM for high-precision bug localization in Verilog is feasible, which is a promising research direction worth further exploration.

The whole workflow of this research is shown in Fig. \ref{fig:workflow}.

%% file: method.tex
This section presents the training process of LiK, which includes three steps: Firstly, we use data from the open-source community and synthetic data to conduct two rounds of unsupervised continual pre-training to enhance the base model's basic knowledge on Verilog bug localization. Then, to enable the model to output the correct bug localization, we fine-tune the model with bug localization samples generated from common Verilog error patterns. Finally, to reduce the errors made by the model, we further finetune it through SimPO algorithm.
\subsection{Continual Pre-Training Using Verilog/C Corpus and Bug Localization Knowledge}

This step involves two rounds of continual pre-training on the base model. The first round is to enhance the model's understanding on basic Verilog knowledge, while the second is target at the task of functional bug localization. In order to reduce training costs while retaining the base model's coding capabilities, we use LoRA algorithm \cite{LoRA} for pretraining, in which only a small portion (0.9142\% in our pretraining process) of all the model parameters are updated. 
\subsubsection{Pretraining with basic Verilog Knowledge}
In the first round of pretraining, we use \xi{Veriseek} \cite{Veriseek} dataset, which consists of Verilog and C corpora as described below:

(a) Verilog corpus. This part of data is from \xi{VGen} \cite{DATE-23-2} and includes two parts: (1) Open-source Verilog files from GitHub, in which each data entry contains "module" and "endmodule" statements and does not exceed 20000 characters in length; (2) Corpus from the e-PDF of 70 Verilog textbooks, where irrelevant content is filtered out and only entries containing Verilog code are retained.

(b) C corpus. According to \cite{ICML-24}, since the syntax of the C language is similar to Verilog, incorporating some C corpus into the model training can enhance its understanding of Verilog syntax. This part of corpus is extracted from the C language documentation in CodeSearchNet \cite{Cdataset} dataset.  

We use MinHash and LSH algorithms to further deduplicate the Veriseek dataset. The final dataset contains a total of 91787 entries, with a total size of 206.48MB, comprising 171.03MB of Verilog-related corpus and 35.45MB of C corpus.

\subsubsection{Pre-Training with Bug Localization Reasoning Thoughts}\label{sec:pt2}
Regarding our research problem, we also need to further pretrain the model with samples of bug localization. 
\begin{figure}[h]
	\centering
	\includegraphics[width=8cm]{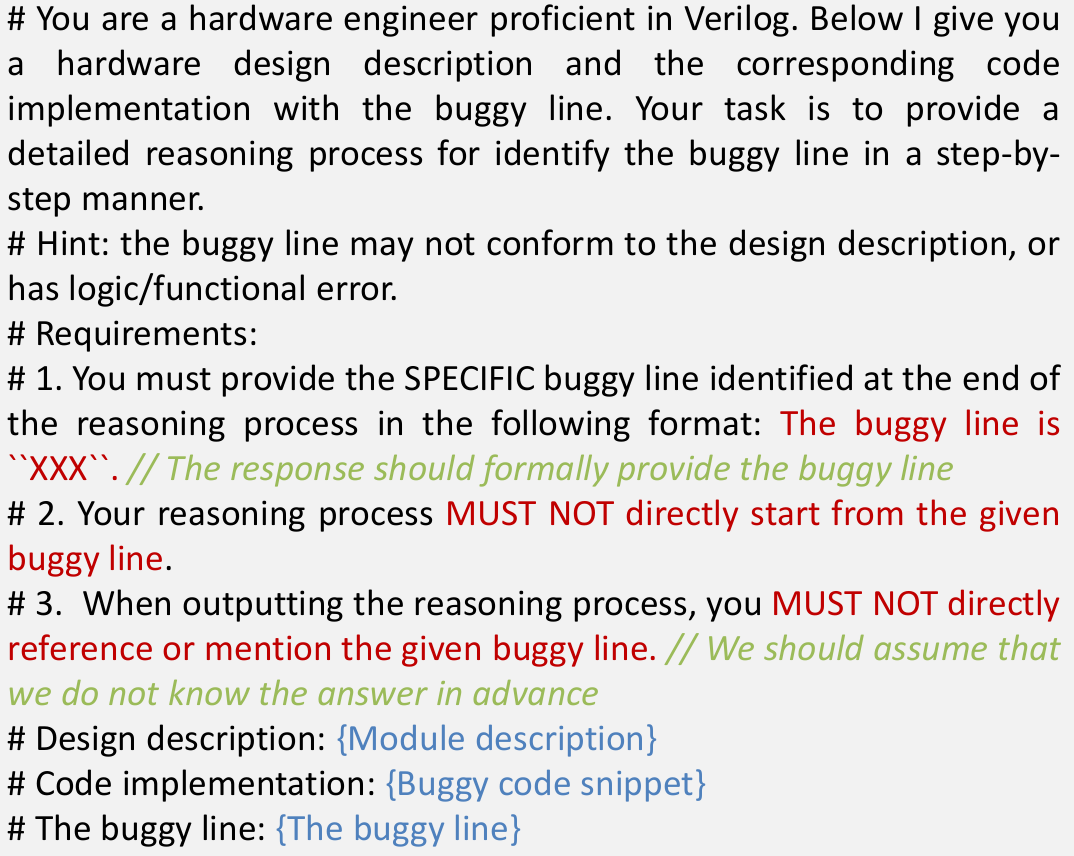}
	\caption{The prompt for reasoning thought generation.}
	\label{fig:fanxiang_prompt}
\end{figure}
According to \cite{CoT}, using reasoning thoughts that lead to correct the answer for pretraining can enhance the LLM's understanding of input tasks, enabling it to generate more relevant answers. Hence, we apply reasoning thoughts as training samples for this round of pretraining, and each data sample is formatted as (\xi{f} $\circ$ \xi{d} $\circ$ \xi{b} $\circ$ \xi{r}), in which $\circ$ denotes the concatenate operator, and \xi{f}, \xi{d}, \xi{b}, \xi{r} represents the sentence ``Find the buggy line in the Verilog code", the design description, the buggy code and the reasoning process respectively. Next we discuss the generation of these training samples.

(a) Seed dataset. We use a combination of three Verilog datasets as the seed data to generate training samples, including \xi{VGen}, an OpenCores \cite{Opencores} dataset and an internal dataset. Each dataset entry is a Verilog module. We apply the MinHash+LSH algorithm to remove duplicate modules, and retain modules that could be synthesized successfully by Synopsys Design Compiler and have more than 20 lines of code. We then leverage GPT-4 to generate the design description of each model, with the prompt shown in Fig. .

(b) Sample generation. For each sample in the seed dataset, we first inject bugs into it and then generate the reasoning thought for locating the buggy line by reverse engineering. Next we discuss this step in detail. 

(b-1) Bug injection. According to \cite{bugpattern}, we consider the following five common bug patterns in Verilog:
\begin{itemize}
	\item Misuse of operators. e.g., Mistaking ``+" for ``-", ``\&" for ``||".  
	\item Errors in numerical values. For example, bit width error.
	\item Keyword errors. For instance, the misuse of ``wire" and ``reg".
	\item Variable name confusion, such as ``counter1" mistakenly written as ``counter2".
	\item Edge errors. For example, mistaking a rising edge ``posedge" for a failing edge "negedge".
\end{itemize}

For each sample, we select one of the above patterns and use regular expressions to modify a random location in the code, thus obtaining the buggy code and buggy line. Note that since we focus on functional bugs, samples with buggy code causing synthesis errors are filtered out.   

(b-2) Reasoning thought generation. Given the buggy line for each sample, we employ off-the-shelf commercial LLMs to generate the reasoning thought for locating it. When prompting the LLM, naturally, we should assume that we do not know the buggy line before reasoning (e.g., Outputs starting with ``Given the buggy line is XXX" are incorrect). Thus, the thought should not directly mention the buggy line. Besides, we tell the model to formally provide the located line at the end of the thought. This enables the automatic verification of each generated thought. To summarize above, the prompt for thought generation is shown in Fig. \ref{fig:fanxiang_prompt}.

\begin{figure}[h]
	\centering
	\includegraphics[width=8cm]{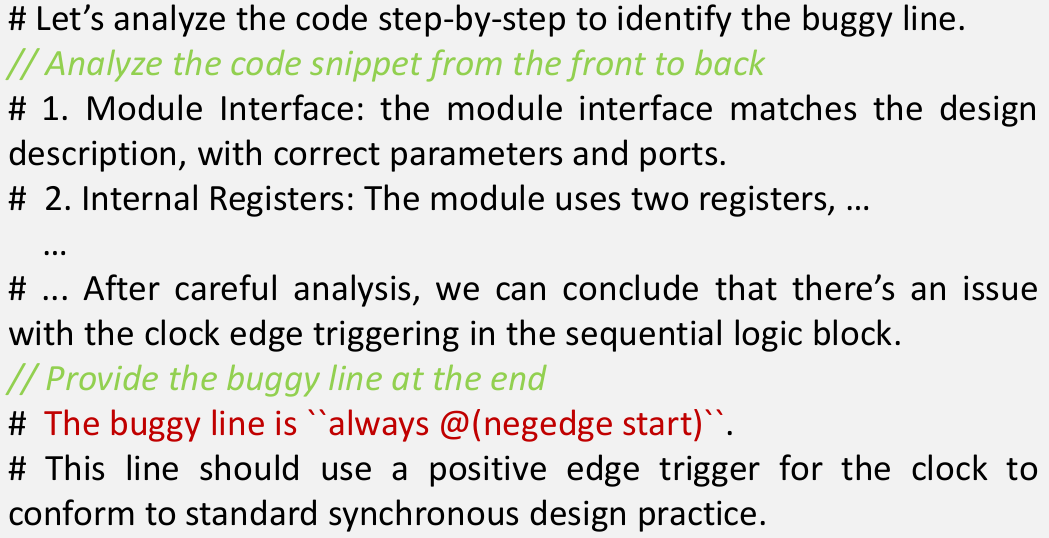}
	\caption{A generated reasoning thought for bug localization.}
	\label{fig:reasoning_thought}
\end{figure}

We use Claude-3.5-Sonnet \cite{Claude} for thought generation and randomly checked 100 of them manually, and we observe that all thoughts follows a human-like reasoning process, it analyzes each part of the input code snippet from front to back and finally provided the buggy line. An example is shown in Fig. \ref{fig:reasoning_thought}. After filtering out the thoughts that do not identify the buggy line correctly, we ultimately obtained 7622 data samples with a total size of 46.83MB. 

\subsection{Supervised Fine-tuning Using Bug Localization Samples}\label{sec:sft}
After pretraining, the model acquires the fundamental knowledge for functional bug localization. However, since pretraining is unsupervised, the model still does not know how to answer the question ``Where is the bug?". To solve this problem, we use samples of bug localization to perform supervised finetuning on the model. We use the LoRA algorithm to finetune the model under the supervision of bug localization samples.

\begin{figure}[h]
	\centering
	\includegraphics[width=8cm]{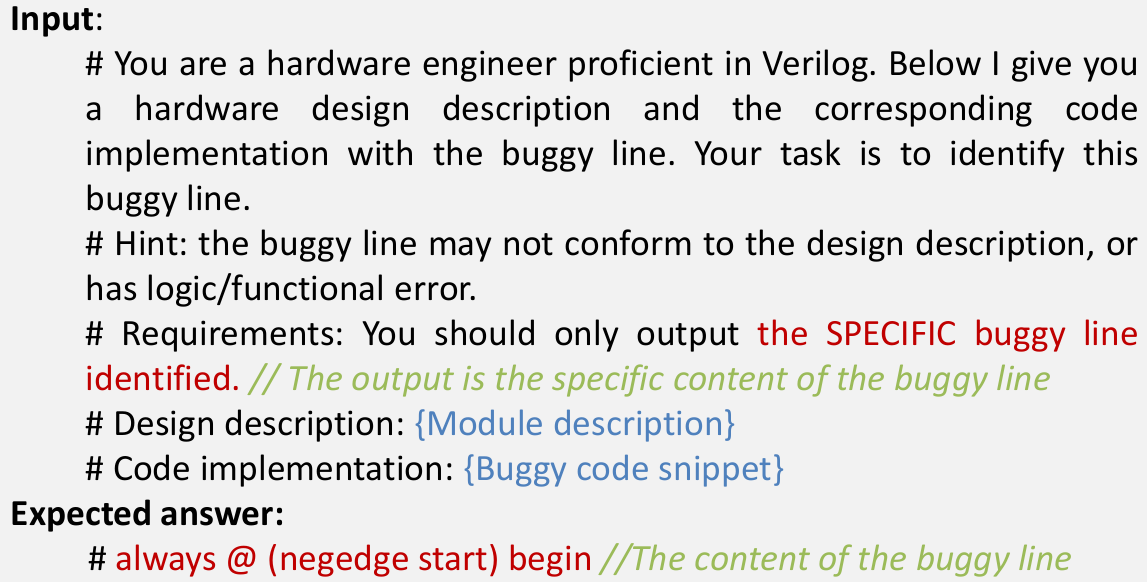}
	\caption{A training sample for supervised finetuning.}
	\label{fig:sft_data}
\end{figure}

In terms of the training data, each of the training sample could be represented as (\xi{p}, \xi{l}), in which \xi{p} is the input to the model and \xi{l} is the expected answer. The input \xi{p} should be the natural language description of bug localization problem, including the module design description and the buggy code. As for the answer \xi{l}, A natural idea is to directly apply the buggy line number, just like existing non-LLM solutions. However, LLMs are generally trained on natural language text, with relatively little data on mathematical reasoning. This indicates that if we use numbers as the output, the answer accuracy is likely to decrease. Therefore, we should have the model output the specific content of the buggy line, rather than its line number. A training sample is shown in Fig. \ref{fig:sft_data}.

We use the same seed dataset and method as in Sec. \ref{sec:pt2} for training data generation, producing a total of 7,866 data samples with a size of 36.72MB. Note that we do not use the reasoning thoughts as the expected output of the model, and this issue will be discussed in Sec. \ref{sec:discuss}.

\subsection{Reduce the Error using SimPO}\label{sec:simpo}
The supervised finetuning in the last subsection teaches the model ``What answers are good". In this step, we further finetune the model using Reinforcement Learning with Human Feedback (RLHF \cite{RLHF}) to let the model know "What answers are bad", thereby reducing the response error. Specifically, we use the SimPO algorithm \cite{SimPO}. In each episode of SimPO, the model receives input from training samples and produces output. It then calculates the reward for the output and objective function based on the preference data, and uses gradient descent to maximize the objective function. The details are as follows.
\subsubsection{Training data of SimPO} 
According to \cite{SimPO}, each training sample could be represented as triples (\xi{x}, \xi{y$_w$}, \xi{y$_l$}), in which \xi{x} is the input, \xi{y$_w$} and \xi{y$_l$} are the positive and negative examples of the output that reflecting human preferences. We use the same seed dataset and method as in Sec. \ref{sec:sft} to generate training samples, where \xi{x} is the natural language description for bug localization problem and \xi{y$_w$} denotes the buggy line. As for the negative example \xi{y$_l$}, to reduce output errors, we set \xi{y$_l$} as a correct line in the buggy code snippet that the model may identify as the buggy line. To be specific, we input \xi{x} into the model and obtain multiple outputs. Then, set \xi{y$_l$} as the line that appears most frequently among all the non-buggy lines output by the model. If all the outputs are buggy lines, randomly pick a line in the buggy code snippet as \xi{y$_l$}. 

We generate 3,542 training samples, with a total size of 19.82MB.
\subsubsection{Reward and Objective Function} SimPO directly aligns the reward function with the generation process of the LLM. Specifically, consider an LLM $\pi_\theta$, which receives input \xi{x} and outputs \xi{y}. The following formula could represent the probability of $\pi_\theta$ generating \xi{y} given \xi{x} as the input:
\begin{equation}
	p_{\theta}(\textit{y} \mid \textit{x}) = \frac{1}{|\textit{y}|} \log \pi_{\theta}(\textit{y} \mid \textit{x}) = \frac{1}{|\textit{y}|} \sum_{i=1}^{|\textit{y}|} \log \pi_{\theta}(\textit{y}_i \mid \textit{x}, \textit{y}_{<i})
\end{equation}

Given $\beta$ as the constant controlling the magnitude of reward differences, SimPO uses $\beta$$\times$$p_{\theta}(\textit{y} \mid \textit{x})$ as the reward of output \xi{y} given \xi{x} as the input. Then, SimPO applies the Bradley-Terry objective function, which represents the difference between the rewards of \xi{y$_w$} and \xi{y$_l$}. To enhance \xi{y$_w$}'s advantage during training, SimPO introduces a difference term $\gamma$ in the function, ensuring that \xi{y$_w$}'s reward exceeds \xi{y$_l$} by at least $\gamma$. Thus, the objective function is:
\begin{equation}
	\mathcal{L}(\pi_\theta) = \mathbb{E}\left[ \log \sigma \left(\beta\times{p_{\theta}}(\textit{y$_w$} \mid \textit{x}) - \beta\times{p_{\theta}}(\textit{y$_l$} \mid \textit{x}) - \gamma \right) \right]
\end{equation}

It is not hard to see that by maximizing the above function, the model is more likely to output the buggy line \xi{y$_w$} instead of \xi{y$_l$}, thus reducing errors tend to be made by the model.

%% file: evaluation.tex
\subsection{Experimental Settings}
\subsubsection{Training Setups}
We trained LiK based on an open-source coding LLM Deepseek-coder-V2-Lite-base-16B \cite{deepseekcoder} using 8 A800-80G GPUs. We apply the cosine annealing to adjust the learning rate during the training process, with warmup steps accounting for 10\% of all training steps. The training hyperparameters are shown in Table. \ref{parameter}. Two rounds of pre-training, supervised fine-tuning, and SimPO training take approximately 1, 3 and 20 hours respectively.
\begin{table}[h]
	\caption{Hyperparameters for Model Training.}
	\label{parameter}
	\begin{tabular}{ccccc}
		\toprule
		Hyperparameter& PT-R1 & PT-R2 & SFT & SimPO\\
		\midrule
		Training batch size & 24 & 32 & 168 & 64\\
		Learning rate & 1e-4 & 1e-4 & 1e-4 & 1e-5\\
		Cutoff length (tokens) & 4096 & 2048 & 2600 & 2600\\
		Number of training epochs& 0.2 & 0.4 & 3.0 & 10.0\\
		$\beta$ & N/A & N/A & N/A & 2.0 \\
		$\gamma$ & N/A & N/A & N/A & 1.0\\
		\bottomrule
	\end{tabular}
\end{table}
\subsubsection{Evaluation Metrics}
We apply the \xi{pass@k} metric, which is commonly used to evaluate the coding abilities of language models \cite{passrate}. It refers to the probability that at least one out of \xi{k} generated answers to a given problem is correct. According to \cite{passrate}, the following formula yields an unbiased estimate of \xi{pass@k}: 
\begin{equation}
\xi{pass@k} := \mathbb{E}\left[ 1 - \frac{\binom{n-c}{k}}{\binom{n}{k}} \right]
\end{equation}
in which \xi{n}\textgreater\xi{k} is the number of generated answers for each test case. In our experiment, we set \xi{n} = 20 and measure \xi{pass@1} and \xi{pass@5}. Obviously, \xi{pass@1} reflects the accuracy of model's single response, while \xi{pass@5} evaluates the performance when multiple answers are generated for each testcase.

In the evaluation of LiK, given the output of the model, the line within the buggy code snippet that has the smallest edit distance from the output is identified as the line located by LiK. 
\subsubsection{Testset} 
The testset is constructed based on RTLLM \cite{RTLLM} dataset, which is commonly used for evaluation on Verilog coding tasks using LLMs. RTLLM consists of 29 typical hardware module design, each of which includes three parts: (1) The design description, which only specifies the design requirements and the input/output interfaces, without mentioning the internal implementation logic; (2) The golden code, with the length ranging from 35 to 175 lines; (3) Testbench. For our testset, we insert bugs into the golden code of RTLLM based on common Verilog error types mentioned in Sec. \ref{sec:pt2}, ensuring that the resulted buggy code does not cause synthesis errors but fails the testbench tests. Each testcase is formatted as the input shown in Fig. \ref{fig:sft_data}. In total, we obtain 102 testcases, with the number of each bug type shown below:
\begin{table}[h]
	\label{errortype}
	\begin{tabular}{@{\hspace{4pt}} c @{\hspace{4pt}} c @{\hspace{4pt}} c @{\hspace{4pt}} c @{\hspace{4pt}} c @{\hspace{4pt}} c @{\hspace{4pt}}}
		\toprule
		 Bug type& Operators & Numerical & Variable &  Keyword & Edge\\
		\midrule
		Number& 25 & 26 & 24 & 13 & 14\\
		\bottomrule
	\end{tabular}
\end{table}

There are few bugs related with keyword and edge in our testset. As for the edge errors, this is because some cases of RTLLM are purely arithmetic logic without edge-triggered mechanisms. Regarding keyword errors, the misuse of some common keywords often causes synthesis errors (e.g., blocking assignment "=" and non-blocking assignment "<="), or being able to pass the testbench tests (e.g., "wire" and "reg"), which do not meet our requirements for testcases. 

\subsubsection{Baselines}\label{baseline} 
For comparison, we employ GPT-3.5, GPT-4 and Claude-3.5-Sonnet as LLM-related baselines. In the evaluation, we find that responses from GPT and Claude often includes multiple code lines and some reasoning process, and the output formats are quite diverse. This makes it difficult to automatically extract the located line from the response. Thus, we consider their answer to be correct as long as the buggy line appeared in their outputs. 

As for the non-LLM baseline, we evaluate Strider\cite{Strider}, which is the SOTA bug localization algorithm for Verilog. Based on testbench, Strider outputs a list of possible buggy lines by comparing the output waveform of the buggy code snippet with that of the golden code. To evaluate Strider on our testset, we modify the testbench provided in RTLLM to derive the correct output waveforms. Also, since the output of Strider is fixed for a given input, we evaluate its hit rate, which is the proportion of test cases where the buggy line is within its output list. 

\subsection{Experimental Results}
\subsubsection{Main results}
We evaluate LiK after each training stages, with the model's temperature set to 0.3. As shown in Table \ref{performance}, the base model of LiK, Deepseek-coder-V2-Lite-16B, does not perform well. This is because there is relatively little data related to Verilog and bug localization in its training corpus. As the training progresses, the performance of LiK continues to increase. Particularly, after the continual pretraining with the reasoning thought for bug localization, the model's performance improves compared with that using only Verilog and C corpus (i.e., the localization accuracy of LiK$_{PTwt}$ and LiK$_{PTwt+sft}$ is higher than that of LiK$_{PTwot}$ and LiK$_{PTwot+sft}$ respectively). This indicates that to enable the model to perform bug localization, during the pretraining stage, it is not sufficient to use only the code data; it is also necessary to use corpus specialized to the task of bug localization. In addition, SimPO training reduces the error made by the model, further improving the LiK's performance. 
\begin{table}[h]
	\centering
	\vspace{-10pt}
	\begin{tabular}{lcccc}
		\toprule
		Type & Method & pass@1 & pass@5 & hit rate\\
		\midrule
		\multirow{3}{*}{\makecell{LLM-\\related}} & GPT-3.5 & 42.4 & 70.2 & N/A\\
		& GPT-4  & 77.9 & 93.9 & N/A \\
		& Claude-3.5 & 90.8 & 93.11 & N/A \\
		\midrule
		\makecell{non-\\LLM} & Strider\cite{Strider} & N/A & N/A & \makecell{58.8+ \\ (19.6?) \\ = 78.4} \\
		\midrule
		Base & \begin{tabular}[c]{@{}l@{}}Deepseek-Coder-\\V2-Lite-16B\end{tabular} & 13.43 & 19.59 & N/A \\
		\midrule
		\multirow{3}{*}{Ours} & LiK$_{PTwot}$ & 18.97 & 27.01 & N/A \\
		& LiK$_{PTwt}$ & 29.15 & 38.76 & N/A \\
		& LiK$_{PTwot+sft}$ & 86.17 & 87.23 & N/A \\
		& LiK$_{PTwt+sft}$ & 87.54 & 90.83 & N/A \\
		& LiK$_{PTwt+sft+simpo}$ & \cu{93.38} & \cu{94.10} & N/A\\
		\bottomrule
	\end{tabular}
	\vspace{0.5cm}
	\caption{Evaluation results. "\xi{PT$_{wot}$}" and "\xi{PT$_{wt}$}" denotes the pretraining process without/with the reasoning thought for bug localization (i.e., The second round of pretraining).}
	\label{performance}
	\vspace{-15pt}
\end{table}

Compared to LLM-related baselines, the performance of LiK surpasses GPT-4 and is comparable to that of Claude-3.5-Sonnet. Note that, as mentioned in Sec. \ref{baseline}, the appearance of the buggy line in the response of GPT and Claude does not necessarily mean that the line located actually located by them is the buggy line. Thus, the experimental data is in fact better than the true performance of GPT-3.5, 4 and Claude. 

As for Strider, for 20 testcases, it could not return a list of buggy lines, which accounts for 19.6\% of all the testcases. Hence, the hit rate of Strider is at most 78.4\%. Mention that in our evaluation, the list computed by Strider includes 6.67 lines on average, while LiK always returns only one line. Overall, starting from all mismatched output signals, Strider identifies some nodes related to these signals in the abstract syntax tree of the code, with each node corresponding to one possible buggy line. Theoretically, this method dictates that the output of Strider comprises multiple lines. In comparison, LiK could directly pinpoint the buggy line with high precision without the need for testbenches, assertions or other EDA tools. This offers a promising new approach for Verilog bug localization. 
\subsubsection{Result on Bug Repair Using LLMs}\label{sec:bugrepair} 
Fundamentally, the task of bug localization is to facilitate the code repair process. This part briefly examines the impact of bug locations, as computed by LiK, on the process of automatic code repair using LLMs. Specifically, we conducted experiments using GPT-3.5 on the same testset for bug localization. We use a simple prompt to instruct the model to output the fixed code, which contains the design description, the buggy code and the buggy line located by LiK. A repair is considered successful if the fixed code passes the testbench without syntax errors. For comparison, we consider the scenario where the located buggy line is removed from the prompt. 

The experimental result shows by adding the buggy line located by LiK into the prompt, \xi{pass@1} and \xi{pass@5} increases from 40.39\% and 63.77\% to 58.92\% and 80.89\% respectively. This significant improvement highlights the critical importance of bug localization data in LLM-assisted Verilog code repair. 

%% file: discussion.tex
\begin{figure}[t]
	\centering
	\begin{minipage}{4.2cm}
		\centering
		\includegraphics[width=4.2cm]{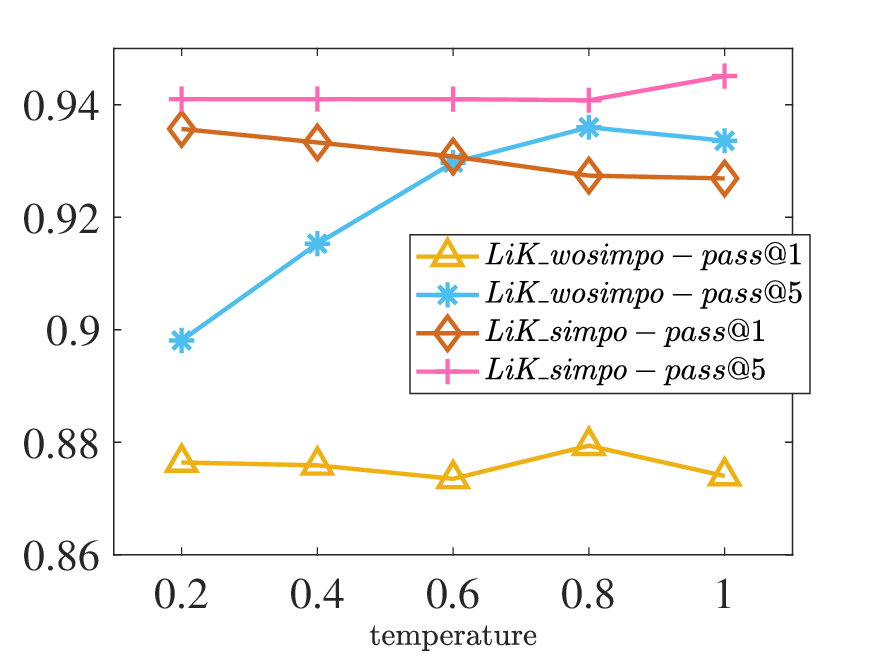}
		\caption{Performance vs temperature.}
		\label{fig:temperature}
	\end{minipage}%
	\hfill
	\begin{minipage}{4.2cm}
		\centering
		\includegraphics[width=4.2cm]{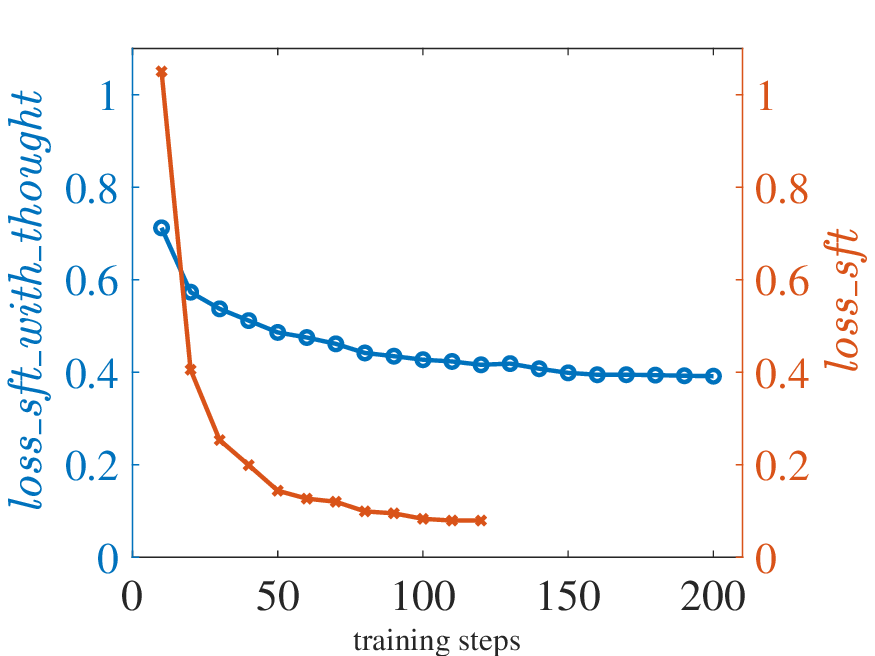}
		\caption{The training loss of SFT with different types of data.}
		\label{fig:loss}
	\end{minipage}
\end{figure}

\subsection{The Impact of SimPO Training}
Sec. \ref{sec:simpo} highlights that SimPO training could reduce common errors in model responses. To validate this, we compare the performance of LiK before and after SimPO training. The performance is evaluated as a function of model temperature, which is typically set between 0 and 1. A higher temperature leads to more varied responses. 

Fig. \ref{fig:temperature} shows that prior to SimPO training, \xi{pass@5} score varies significantly with the temperature, and is notably higher than \xi{pass@1}. This indicates that the model's output for the same testcase is very unstable, leading to more incorrect answers. In SimPO training, we use wrong model outputs as the negative examples of each training sample, which enhances the stability of the response and thereby reduces common errors by the model. This is illustrated in Fig. \ref{fig:temperature}: through SimPO training, the gap between \xi{pass@5} and \xi{pass@1} significantly decreases, and \xi{pass@5} becomes more stable with changes in temperature. In addition, under all temperature settings, LiK trained with SimPO has higher \xi{pass@1} and \xi{pass@5} scores. This demonstrates that SimPO training provides benefits for both scenarios requiring high-precision answers and those allowing multiple attempts.   

\subsection{Why not Using Reasoning Thought for SFT?}
In the second round of continual pretraining, we use the reasoning thoughts for bug localization as the training corpus. Moreover, as mentioned in \cite{CoT}, prompting the LLM to output the reasoning thought instead of directly providing results could improve the answer accuracy. Thus, a natural idea is to let the model output the reasoning thought, as shown in Fig. \ref{fig:reasoning_thought}. For this purpose, in the SFT phase, we need to use the reasoning thought rather than just the buggy line as the label of each training sample. However, intuitively, reasoning thoughts are not the intended output. Introducing them during SFT could potentially introduce extra noise, leading to decreased accuracy. To verify this, we finetune the pretrained model using both reasoning thoughts (Generated as in Sec. \ref{sec:pt2}) and the buggy lines (What we currently do in Sec. \ref{sec:sft}), while tracking the loss changes during training.

As shown in Fig. \ref{fig:loss}, the model trained with reasoning thoughts exhibits significantly higher training loss compared with that trained with only the answers (i.e., \xi{loss\_sft\_with\_thought} vs \xi{loss\_sft}). This indicates that reasoning thoughts bring additional noise into the training process, preventing the model from fitting the training data well. This also leads to a decrease in answer accuracy: when temperature equals to 0.3, the model trained with reasoning thoughts achieve \xi{pass@1} and \xi{pass@5} score of only 73.48\% and 77.69\% respectively, which are lower than 87.54\% and 90.83\% by the model trained directly with the answer. Hence, we use the localization results directly as data labels for SFT phase.

%% file: conclusion.tex
This paper proposes LiK, an LLM-based solution for locating the functional bugs in Verilog codes. LiK is trained through two rounds of continuous pretraining with Verilog/C corpora and reasoning thoughts for bug localization, supervised finetuning with bug localization samples and SimPO algorithm with human preference data. Experimental results show that LiK achieves high localization accuracy, which surpasses GPT-4 and is comparable to Claude-3.5-Sonnet. Also, by utilizing LiK's outputs, the debug efficiency of GPT-3.5 is significantly improved. To our knowledge, this is the first work to consider LLM-based Verilog bug localization. The model and materials of this research will be open-sourced later.   

This research also indicates that using LLMs for Verilog bug localization is a promising topic, which can be further explored in following aspects:

(1) Leveraging more information. Currently, LiK only takes the design description and the code snippet as its input. If more expert information or feedback from EDA tools (e.g., Mismatched waveforms from testbench simulation) could be utilized, the performance of LiK might be further improved.

(2) Extending to more general scenarios. Since existing works mainly considered the problem within a single module, we will extend LiK for more complex hardware designs (e.g., a 5-stage pipeline). Additionally, we aim to adapt LiK for other hardware design languages, including VHDL, SystemVerilog, and Chisel. 

(3) Integration with other components for LLM-based debugging. As mentioned in Sec. \ref{sec:bugrepair}, the location of bugs is crucial for LLM-assisted Verilog debugging. Thus, we will develop a Verilog debugging framework consists of multiple LLM agents, each dedicated to a specific task such as bug localization and repair.